\documentclass[aps,prb,superscriptaddress,showpacs,secnumarabic, eqsecnum,twocolumn, floatfix]{revtex4}
\usepackage{graphicx}
\usepackage{dcolumn}
\usepackage{bm}
\usepackage{pstricks}

\newcommand{\ba}{\begin{eqnarray}}
\newcommand{\ea}{\end{eqnarray}}
\def\be{\begin{equation}}
\def\ee{\end{equation}}

\begin{document}
\title{Fermi Liquid instabilities in two-dimensional lattice models}
\author{C.A. Lamas}
\affiliation{Departamento de F\'isica, Universidad Nacional de La
Plata, C.C. 67, 1900 La Plata, Argentina}
\author{D.C. Cabra}
\affiliation{Laboratoire de Physique Th\'{e}orique, Universit\'{e}
Louis Pasteur, 3 Rue de l'Universit\'{e}, 67084 Strasbourg, C\'edex,
France.} \affiliation{Departamento de F\'isica, Universidad Nacional
de La Plata, C.C. 67, 1900 La Plata, Argentina}
 \affiliation{Facultad de Ingenier\'\i a, Universidad
Nacional de Lomas de Zamora, Cno.\ de Cintura y Juan XXIII, (1832)
Lomas de Zamora, Argentina.}
\author{N. Grandi}
\affiliation{Departamento de F\'isica, Universidad Nacional de La
Plata, C.C. 67, 1900 La Plata, Argentina}
%

\begin{abstract}
We develop a procedure for detecting Fermi liquid instabilities by extending the analysis of Pomeranchuk to two-dimensional lattice systems. The method is very general and straightforward to apply, thus providing a powerful tool for the search of exotic phases. We test it by applying it to a lattice electron model with interactions leading to $s$ and $d$-wave instabilities.

\end{abstract}


\maketitle
\section{Introduction}

The Landau theory of the Fermi Liquid (FL) is one of the most
important frameworks to understand conventional weakly
interacting metallic systems \cite{review leggett}. The low energy
physics of interacting fermions in three dimensions is usually
described by Landau's FL theory whose central assumption is the
existence of single particle fermionic excitations, or
``quasiparticles", with a long lifetime at very low energies. In
lower dimensions, however, the situation is much more interesting:
in one dimensional systems Landau's quasiparticles are typically
unstable, giving rise to the so called Luttinger Liquid. On the
other hand, two dimensional lattice models are far more complicated
to treat since conventional perturbation theory breaks down for
densities close to half-filling, where competing infrared
divergences appear as a consequence of Fermi Surface (FS) nesting
and van Hove singularities.

In Ref. \cite{Pomeranchuk} Pomeranchuk developed a method to
diagnose instabilities of the FL, by ``deforming'' the FS and studying
the resulting energy gain. In its original form, it applies to
systems with a three dimensional spherical FS, but it can be easily
generalized to the two-dimensional continuum case.

The experimental observation of exotic phases in strongly correlated
systems has triggered an enormous effort from the theoretical
community to try to understand their microscopical origin. One
possible route to detect instabilities of a FL is precisely the
analysis done by Pomeranchuk. Due to that, the Pomeranchuk
instability has been studied by several authors with different
techniques in the last few years
\cite{Metzner1,Metzner2,Metzner3,Quintanilla1,Quintanilla2,Wu,Nilson}
and in particular, the instability of the FL towards the nematic
phase was investigated for several models
\cite{Metzner_PRL,nematic_MF,Hankevych,Kampf,Kee1,Kee2}.

In this paper we develop a general method to trace such
instabilities in lattice models, in a simple and rigorous way. It allows for the study of systems which have an arbitrary shape
of the FS in the absence of interactions, thus being applicable to
models relevant to high temperature superconductors, manganites,
ruthenates, etc. \cite{Grigera1,Grigera2,Kee3,Yamase}. as long as
one can rely on a perturbative analysis. It can be applied in
principle to any lattice problem in a systematic way.

We test our method within two examples, the attractive Hubbard model
and a model with forward scattering interactions that give rise to $d$-wave FS deformation (the so-called ``$d$-wave
Pomeranchuk instability").

The paper is organized as follows: Section \ref{sec:derivation}
contains a detailed derivation of the method, with the proof of our
formulas in subsection \ref{sec:proof}, and a shorthand recipe for
the application of the results in subsection \ref{sec:recipe}. Then
in Section \ref{sec:Hubbard} we apply the method to a two
dimensional square lattice with the various interactions studied in
\cite{Metzner_PRL}, the $s$-wave interaction being studied in
subsection \ref{sec:s-wave}, while the $d$-wave instability in
subsection \ref{sec:d-wave}. Finally, section \ref{sec:conclusions}
contain the conclusions, and some specific calculations are
presented in the Appendix.

\section{Two dimensional Pomeranchuk Instability}
\label{sec:derivation}
\subsection{Derivation of the method}
\label{sec:proof}
In the theory of Landau's FL, the free dynamics at zero temperature is determined by the
dispersion relation $\varepsilon(k)$. In terms of it, the FS is defined as the set of
points in momentum space
satisfying the equation
\be
\mu-\varepsilon(k)=0\, .
\ee
In the ground state of the system, all single-particle states inside the FS $\mu-\varepsilon(k)>0$ are occupied, while those outside FS  $\mu-\varepsilon(k)<0$ are not. Excited states of the system are built by moving some particles from the inner single-particle states to the outer ones.

The energy of such excited state as a functional of the change in
the equilibrium distribution function can be written as
\small
\ba
&&\!\!\!\!\!\!\!\!
E\!=\!\int \!\!d^{2}\!k\,(\varepsilon(k)\!-\!\mu)\delta n(k) 
+
\frac{1}{2}\!\int \!\!d^{2}\!k \!\!\int \!\!d^{2}\!k'
f(k,k')\;\delta n(k)\delta n(k')\, ,
\nonumber \\
&&~\!\!\!\!\!\!\!\!
\label{deltaE} \ea \normalsize
where $\delta n(k)$ is the change in the distribution function
$n(k)$, and we have assumed that only two-particle interactions are present. The interaction function $f(k,k')$ can be related to the low
energy limit of the two particle vertex.

Pomeranchuk criterion allows to identify low energy excited
states of the system that make (\ref{deltaE}) negative. This signals
an instability, and the breakdown of the FL description. In what
follows, we will carefully go through all the steps needed to
perform such analysis.

~

First let us define, associated to any given state of the system, a smooth function $g(k)$ such that it takes positive values at occupied single-particle states and negative values at unoccupied ones. Then at the frontier between these two regions we have the equation
\be
g(k)=0\, .
\label{UnperturbedFS}
\ee
For the ground state, such frontier coincides with the FS allowing us to choose
\be
g(k)=\mu-\varepsilon(k)\, . \label{uno}
\ee

Under a variation $\delta n(k)$ of the distribution function $n(k)$,
we get an excited state that can be described in terms of a new function
$g'(k)=g(k)+\delta g(k)$. The frontier between occupied and unoccupied single-particle states is now located at points satisfying
\be g'(k)=g(k)+\delta g(k)=0\, . \ee
By an abuse of language we will call the solution of this equation the {\em deformed FS}.

Since at $T=0$ we have $\delta n(k)=\pm1$ we can write
\ba
 \delta n(k)&=&H[g'(k)]-H[g(k)]\, ,
 \label{unoo}
\ea
where $H(x)$ is the unit step function, defined by $H(x)=1$ if $x>0$
and $H(x)=0$ if $x<0$. This can be replaced in (\ref{deltaE}) to
write the energy of the quasiparticles as a functional of $g(k)$ and
$g'(k)$, namely
\begin{widetext}
\ba
E =\int d^2 k \; (\varepsilon(k)-\mu)\left(H[g'(k)]-H[g(k)]\right)
+\frac{1}{2}\int d^2 k \int d^2 k' f(k,k')
(H[g'(k)]-H[g(k)])(H[g'(k')]-H[g(k')]).\ \ \label{deltaEH} \ea
\end{widetext}
To go further, we have to take into account the constraint imposed
by the Luttinger theorem\cite{Luttinger}, or in other words the
preservation of the area of the FS under the deformation
\ba \int d^{2}k \;\delta n(k)\equiv 0\,  .\ea
By using (\ref{unoo}) this can be rewritten as a functional constraint on the functions
$g(k)$ and $g'(k)$
\ba \int d^{2}k \;H[g'(k)]=\int d^{2}k \;H[g(k)]\, .
\label{constrant} \ea

In two-dimensions the constraint (\ref{constrant}) can be easily
solved as follows. We first rename the integration variables on the right hand side to $k'$. Next we assume that a change of variables $k'=k+\delta k(k)$ can take the right hand side into the form of the left hand side. Writing $g'(k)=g(k)+\delta g(k)$ we get two unknown functions to be solved for, namely  $\delta g(k)$ and $\delta k(k)$, together with the equation
\small
\ba \int d^{2}k\, H[g(k)+\delta g(k)]&\!=\!&\int \!d^{2}k\!
\left|1+\partial_{j} \delta k^{i}\right|H[g(k+\delta k(k))]\,,
\nonumber\\
\ea \normalsize
where $i,j\in\{1,2\}$ label the orthogonal directions in momentum space.

A particular class of solutions can then be obtained by solving the
following equations
\ba
\left|1+\partial_{j} \delta k^{i}\right|&=&1
\,,\nonumber
\\
g(k)+\delta g(k)&=&g(k+\delta k(k)) \,.\label{APDD-2} \ea
The first line (\ref{APDD-2}) implies that the change of variables
going from $k'$ to $k$ is an area preserving diffeomorphism. The
second line \label{APDD2} on the other hand, can be interpreted as
saying that the variation $\delta g(k)$ is a translation of $g(k)$
by an amount $\delta k$. We can solve (\ref{APDD-2}) as
\ba \delta k^{i}&=& (e^{\epsilon^{jk}
\partial_{j}\lambda\partial_{k}}-1)k^i\,,\nonumber
\\
\delta g&=& (e^{\epsilon^{ij}
\partial_{i}\lambda\partial_{j}}-1)g\,.\label{APD}
\ea
Where $\lambda$ is a free function parameterizing the deformation.
If we assume that the deformation of the FS is small, then $\delta
g(k)$ is also small and we can  parameterize it in terms of a slowly varying $\lambda$
 \ba
 \delta k^i&\simeq&\epsilon^{ij}\partial_j\lambda\,,\nonumber\\
 \delta g&\simeq&
\epsilon^{ij}\partial_j\lambda\partial_{i}g
\,.
\label{variation.of.g}\ea
In what follows, each specific form of $\lambda$ will characterize an excited state, the sign of the resulting energy will give us information about the instabilities.

~

Now that we have solved the constraint, we go back to
the energy of the quasiparticles (\ref{deltaEH}) and write it in
terms of the free unconstrained variable $\lambda$. To simplify the resulting
expression, we need to change variables to a more
convenient coordinate system in momentum space. We choose a special set of variables
\ba
g&=&g(k_{x},k_{y})\,,\nonumber \\
s&=&s(k_{x},k_{y}) \,,\label{changeofvariables} \ea
where the new variable $g$ varies in the direction transverse to the
unperturbed FS. The variable $s$ varies in the longitudinal
direction tangent to the FS, namely it satisfies
$\partial_is\partial_ig=0$.

Separating the energy (\ref{deltaEH}) into a linear and an interaction term $E=L+I$, we get for the linear part
\ba
L&=&\int d^2 k \; (\varepsilon(k)-\mu)\left(H[g+\delta
g]-H[g] \right)=
\nonumber\\
&=&\int \!ds\,dg \;J(s,g)(\varepsilon(s,g)-\mu)\left(H[g+\delta
g]-H[g] \right)=
\nonumber\\
&=&\int \!ds\!\int_{-\delta g}^0\! dg \;J(s,g)(\varepsilon(s,g)-\mu)\,,
\ea
where $J=|{\partial(k_x,k_y)}/{\partial (g,s)}|$ is the Jacobian of the
transformation (\ref{changeofvariables}). Expanding in powers of the integration variable $g$ around the unperturbed FS $g=0$ we get
\ba
\!L\!\!&=&\!\!\int \!ds\!\int_{-\delta g}^0\! \!dg \;\partial_{\bar g} \big[J(s,{\bar g} )(\varepsilon(s,{\bar g} )-\mu\big)]_{{\bar g} =0}\,g+{\cal O}(\delta g^2)=
\nonumber \\
&=& \!\!\frac12\int \!ds\;[J(s,\bar g)\delta g^2]_{g=g'=0}+{\cal O}(\delta g^3) \,,
\label{formulalarga1}
\ea
where in the second line we have integrated out the variable $g$ and made use of the fact that $(\varepsilon(s,g=0)-\mu\big)=0$.
In order to replace the explicit form of $\delta g$ eq.(\ref{variation.of.g}) in the integrand of (\ref{formulalarga1}) we make use of the identity
\ba \epsilon^{ij}\partial_i g\partial_j \lambda &=&
\epsilon^{ij}(\partial_ig \partial_g g+\partial_i s\partial_s
g)(\partial_j g\partial_g \lambda+\partial_j s\partial_s\lambda)=
\nonumber\\
&=& \epsilon^{ij}\partial_ig \partial_j s\partial_s\lambda\equiv
J^{-1}\partial_s\lambda\,,
\ea
where we have used the fact that, according to our definitions,
$\partial_gg=1$ and $\partial_sg=0$.
Now replacing in (\ref{formulalarga1}) we get
 \ba
 L&=&\frac{1}{2}\int ds\left[J^{-1}(g,s)\,(\partial_{s}\lambda)^{2}\right]_{g=0}
 \ea
The calculus of $I$ is analogous and gives
\small
 \ba
 I=\frac{1}{2}\int ds\int ds'\; \left.[f(g,s;g',s')\; (\partial_{s}\lambda)(\partial_{s'}\lambda)]\right|_{g=g'=0}.
 \ea
 \normalsize
Adding the two contributions we finally have
\small
\ba
E\!&=&\!\frac{1}{2}\!\int \!ds\int \!ds' \left(
\phantom{\frac12}\!\!\!\! f(0,s;0,s')+J^{-1}(0,s)\delta(s-s')
\right) \times\nonumber\\&&\ \ \times
\partial_{s}\lambda(0,s)
\partial_{\!s'}\!\lambda(0,s')\,.
\label{bilinear0}
\ea \normalsize
As the functions $\lambda(0,s)$ characterizing the excited states are arbitrary, we can equally work
with functions $\psi(s)=\partial_{s}\lambda(g,s)|_{g=0}$. In what follows we will
be interested in excited states such that $\psi(s)\in L_{2}[0,S]$. Assuming that $s$ makes a complete
turn around the FS when it runs from $0$ to $S$, we also need to impose periodicity in that interval.

Since the sign of $E$ in eq. (\ref{bilinear0}) determines the
stability of the FL, from all the above we conclude that the
stability condition reads
\begin{widetext}
\vskip-.3cm
\ba E=\int \!ds'\!\int \!ds \,\psi(s')\frac12
\left(\phantom{\frac12}\!\!\!\! J^{-1}(s)\delta(s-s') + f(s,s')
\right)\psi(s)
>0\;,
\label{bilinear} \ea
\end{widetext}
where have we defined
\ba
 f(s,s')&=&f(g,s;g',s')\big|_{g=g'=0}\,,\nonumber\\
 J^{-1}(s)&=&J^{-1}(g,s)\big|_{g=0}\,.
\ea

Note that the stability condition has two terms, the first of which
contains the information about the form of the FS via $J^{-1}(s)$,
while the second encodes the specific form of the interaction in
$f(s,s')$. There is a clear competition between the interaction
function in the second term of the integrand and the first term that
only depends of the geometry of the unperturbed FS.

We see that $E$ is a bilinear form, acting on the real
functions $\psi(s)$ parameterizing the deformations of the FS
\ba E=\langle\psi,\psi\rangle \,,\ea
where
\small
 \ba\label{eq:pseudo escalar general}
 \nonumber \langle
    u,v\rangle\!\!&=&\!\!
\int \!\!ds'\!\!\!\int \!ds \,u(s)\frac12
\left(\phantom{\frac12}\!\!\!\!
\!f(s,s')+J^{-1}\!(s)\delta(s\!-\!s') \right)v(s')\,.
\\&&
\!\!\!\!\!\!\!\!
\ea
\normalsize %
The stability condition is then equivalent to asking this form to be
positive definite for any possible deformation, {\em i.e.}
\ba \forall\psi: \langle\psi,\psi\rangle>0 \,.\ea
In consequence, the natural way to diagnose an instability is to
diagonalize this bilinear form and to look for negative eigenvalues.

We can expand the functions $\psi(s)$ in some basis of $L^2[0,S]$
that we will denote $\{\xi_{i} (s)\}$
 \ba
 \psi(s)=\sum_{i}\;a_{i}\xi_{i}(s)\,,
 \ea
and then write
\ba
    E=\sum_{i_{1},i_{2}}\;a_{i_{1}}a_{i_{2}}\langle
    \xi_{i_{1}},\xi_{i_{2}}\rangle\,.
\ea

The bilinear form $\langle~,~\rangle$
can be taken as a pseudo-scalar product, which is linear and
symmetric but, in general, not positive-definite. Only in the
free case $f(s,s')\equiv 0$ the positivity is ensured. If
$\{\xi_i(s)\}$ are taken to be orthogonal with respect to this pseudo-scalar product, then the functional
(\ref{bilinear}) is given by
\ba\label{eq:funcional diagonal}
    E=\sum_{i}\;a^{2}_{i}\;\chi^{\mu}_{i}\,,
\ea
where $\chi_{i}^{\mu}=\langle \xi_{i},\xi_{i}\rangle$ is the square
pseudo-norm of the orthogonal functions. If $\chi_{i}^{\mu}$ has a
negative value for some $i$, then by choosing the corresponding $a^{2}_{i}=1$
and $a^{2}_{j}=0$ for $j\neq i$, we see that the energy is negative
denoting an instability. In this case we say that we have an
instability in the $i$-th channel. In other words, the  stability condition
has been mapped into
\ba \forall i: \chi^{\mu}_i>0 \,,\ea
the $\chi^{\mu}_i$ being taken as the stability parameters.
If any of these quantities is negative, then the FS is unstable.

We perform such diagonalization by choosing a basis on $L_2[0,S]$ as
a given set of functions $\{\psi_i\}$ and then making use of the
Gram-Schmidt orthogonalization procedure to transform it into an
orthogonal basis $\{\xi_i\}$. Note that, being the bilinear form not
necessarily positive definite, the new basis cannot be normalized to $1$ but to $\pm 1$.

\vspace{.5cm}

This is our main result. Our method to search for Pomeranchuk instabilities, can be summarized in the following recipe:

\subsection{Recipe}
\label{sec:recipe}

\begin{enumerate}

\item Get the dispersion relation $\varepsilon(k)$ and the interaction
function $f(k,k')$ for the model under study.

\item Change variables according to (\ref{changeofvariables}).
The variable $g$ is completely fixed by the dispersion relation according to  (\ref{uno}).
The choice of $s$ is arbitrary except for the constraint of being
tangential to the FS, $\partial_i s\partial_i g=0$.

\item Write the bilinear form $E$ as in (\ref{bilinear}).

\item Choose an arbitrary basis of functions $\{\psi_i\}$ of $L_2[0,S]$.

\item Apply the Gram-Schmidt orthogonalization procedure, verifying at each
step whether $\langle\xi_i,\xi_i\rangle>0$

\item If for a given channel $i$ one finds that $\langle\xi_i,\xi_i\rangle<0$, the
FS is diagnosed to be unstable.

\end{enumerate}
Note that since $L_2[0,S]$ is infinite dimensional, the present
method is not efficient to verify stability: at any step $i$ it may
always be the case that for some $j$, $\chi^{\mu}_{i+j}<0$. Moreover, we have
not exhausted all the possible solutions of the constraint
(\ref{constrant}) but only explored a subset of them.

\section{Instabilities in the square lattice}
\label{sec:Hubbard}

\subsection{Contribution of the free Hamiltonian}
\label{sec:free}
We start considering free fermions in the square lattice, with a
Hamiltonian given by
 \ba
 H_{0}=\sum (\varepsilon(k)-\mu) c^{\dagger}_{k}c_{k}\,,
 \ea
 where
 \ba
 \varepsilon(k)=-2t(\cos k_{x}+\cos k_{y})\,,
 \ea
where only hopping to nearest neighbors has been taken into account.  The
FS is defined by
 \ba
 g(k)=\mu-\varepsilon(k)=\mu+2t(\cos k_{x}+\cos k_{y})=0\,,
 \ea
where $\mu$ is the chemical potential. Notice that $g>0$ inside the
area bounded by the FS, negative outside it, and zero at the FS.

Now we follow the recipe given in Section \ref{sec:recipe}, changing
variables according to (\ref{changeofvariables})
 \ba
 g(k_{x},k_{y})&=&\mu+2t(\cos k_{x}+\cos k_{y})\,, \\
 s(k_{x},k_{y})&=&\arctan\left(\frac{\tan(k_{y}/2)}{\tan(k_{x}/2)}\right)\,.
 \ea
\begin{figure}[t]
  \includegraphics[width=240pt]{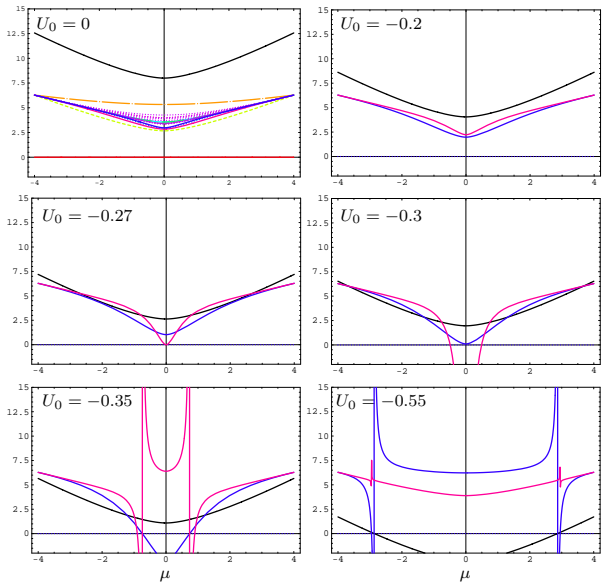}\\
  \caption{The instability parameters. For $U_0=0$ we show the  first $10$ parameters $\chi^{\mu}_{i}$ as a function of $\mu$.
   For other values of the interaction we show only the parameters corresponding to the three lower
   channels
   that show instabilities, namely channels $\chi_{0}$, $\chi_{8}$ and $\chi_{16}$ (colors are identified in Fig \ref{fig:pd_cte}). Notice that when we increase $U_{0}$, the FL breakdown occurs first for the higher channels and closer to half filling.}\label{fig:ordenes_f0}
\end{figure}
$\!\!\!$It is straightforward to see that $g$ and $s$ are mutually
orthogonal variables. Using the following shorthand notation
 \ba \label{alphabeta}
\alpha&=&\cos k_{x}\,,\\
\beta&=&\cos k_{y}\,,
 \ea
we can write
 \ba \label{eq:solutions}
g&=&\mu+2t(\alpha+\beta)\,,\nonumber\\
\tan^{2}(s)&=&\left(\frac{1-\beta}{1+\beta}\right)\left(\frac{1+\alpha}{1-\alpha}\right)\,,
 \ea
and the Jacobian takes the form
 \ba \label{eq:J con alfa}
J=t\left(\frac{\alpha \beta -1}{\alpha^{2}+\beta^{2}-2}\right)\,.
 \ea
Notice that $J\geq0$. Writing $\alpha$ and $\beta$ as functions
$g$ and $s$ we have for the Jacobian evaluated at $g=0$.
\ba
J[g=0,s]=\frac{1}{2\;t\sqrt{1-\beta(\mu)\;\cos^{2}(2s)  } }\,,
\ea
with $\beta(\mu)=1-(\frac{\mu}{4\;t})^{2}$. The limits for the
variable $s$ can be taken as $-\pi< s \leq \pi$.

The inverse of the jacobian $J^{-1}(s)$ can be expanded in series of
$\sin(ns)$ and $\cos(ns)$ and it is straightforward to show that only
the coefficients of $\cos(4ns)$ are non-vanishing. This results in
the following expansion
\ba\label{eq:desarrollo_j}
    J^{-1}(s)=\sum_{n}\;j^{\mu}_{n}\cos(4ns)\,,
 \ea
where the coefficients $j_{n}^{\mu}$  are fixed by the expansion.
Some of them are given in the Appendix. The simplicity of this expansion suggest to take as our starting base in the Gram-Schmidt orthogonalization procedure the set $\{\sin(ns),\cos(ns)\}$.

~

In the next subsections we will analyze as an example of application the
possible instabilities in this two-dimensional fermion model when
subjected to various interactions. In particular we are interested
in interactions of the form \cite{Metzner_PRL}.
 \ba  \label{eq:interaccion_general}
 f(k,k')=\;\text{Constant}\;\times\; d(k)d(k')\,,
 \ea
with
\ba
    d(k)=1 \,, \qquad\qquad\qquad\;\;& \hspace{0.3cm}& \text{($s$-wave)}\,,\\
 \nonumber   d(k)=(\cos k_{x}+\cos k_{y})\,, & \hspace{0.3cm} & \text{(extended $s$-wave)}\,,\\
  \nonumber  d(k)=(\cos k_{x}-\cos k_{y}) \,,& \hspace{0.3cm} & \text{($d$-wave}\;d_{x^{2}-y^{2}})\,.
\ea

%
\subsection{$s$-wave instability}
\label{sec:s-wave}
First we consider  a constant interaction corresponding to
take\cite{Metzner_PRL}
  $d(k)=1$ so that
  \ba
  f(k,k')=U_{0}\,,
 \ea
$U_{0}$ being a constant measuring the strength of the interaction.
This form of the interaction can be obtained by a
Mean Field approximation or a first order perturbative
expansion for the interaction function on the
Hubbard model\cite{Frigeri, Fuseya}.
\begin{figure}[t]
  \includegraphics[width=250pt]{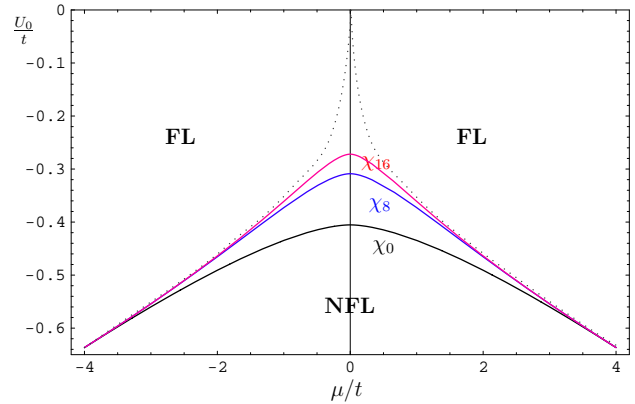}\\
  \caption{Phase diagram for $f(s,s')=U_{0}$, are displayed  the regions of
  instability  for the first three channels. For smaller
   interactions, higher channels are unstable closer to half filling.  }\label{fig:pd_cte}
\end{figure}

Using the product defined in eq. (\ref{eq:pseudo escalar general})
we can calculate the first instability parameters.
\ba \label{eq:chi_for_swave}
  \chi_{0}^{\mu}&=&2\pi( U_{0} \pi+j^{\mu}_{0})\,,\\
  \nonumber  \chi_{1,2}^{\mu}&=&\pi j^{\mu}_{0}\,,\\
  \nonumber  \chi_{3}^{\mu}&=&\pi(j^{\mu}_{0}-\frac{1}{2}j_{1}^{\mu})\,,\\
  \nonumber  \chi_{4}^{\mu}&=&\pi(j^{\mu}_{0}+\frac{1}{2}j_{1}^{\mu})\,,\\
  \nonumber  \chi_{5,6}^{\mu}&=&  -\frac{\pi}{2}
  \frac{{j_{1}^{\mu}}^{2}}{j_{0}^{\mu}}+  j_{0}^{\mu} \left(\pi+\frac{\pi}{4}
  \left(\frac{{j_{1}^{\mu}}}{j_{0}^{\mu}}\right)^{2} \right)\,,\\
  \nonumber
  \chi_{7}^{\mu}&=&\pi(j^{\mu}_{0}-\frac{1}{2}j_{2}^{\mu})\,,\\
  \nonumber
  \chi_{8}^{\mu}&=&\frac{{\pi }^2\,U_{0}\,
     {j_{1}^{\mu}}^2}{2\,
     {\left( \pi \,U_{0} +
         j_{0}^{\mu} \right) }^2} -
  \frac{\pi \,{j_{1}^{\mu}}^2}
   {\pi \,U_{0} + j_{0}^{\mu}}+\\
    &&+ \nonumber
  j_{0}^{\mu}\,
   \left( \pi  + \frac{\pi \,{j_{1}^{\mu}}^2}
      {2\,{\left( \pi \,U_{0} +
            j_{0}^{\mu} \right) }^2} \right)
   + \frac{\pi \,j_{2}^{\mu}}{2}\,,
   \nonumber\\
   &&\!\!\!\!\!\!\!\!\!\!\!\!\!\!\!\!\!\!\text{etc\dots}
\ea

The stability parameters for the first channels are shown in Fig
\ref{fig:ordenes_f0}. For $U_{0}=0$ all the $\chi_{n}^{\mu}$ are
positive as expected. When we increase the interaction, among the
first 20 parameters, only $\chi_{0}^{\mu}$, $\chi_{8}^{\mu}$ and
$\chi_{16}^{\mu}$  change. For simplicity only these
parameters are plotted for $U_{0}\neq 0$.

With these first instability channels we can draw a qualitative
phase diagram in the $(\mu,U_{0})$ space as in Fig
\ref{fig:pd_cte} where the first instability zones are shown and a
tentative global phase diagram is drawn.

Note that, when the interaction is increased, the first instability
channel corresponds to the highest of the three shown in the figure
({\it e.g.} $\chi_{16}^{\mu}$).
This behavior is maintained for channels $\chi_{i}^{\mu}$ with higher index $i$, and we
assume that generically these higher channels will show the instability
closer to half filling and for interactions arbitrarily small. On the other hand, the higher
the channel, the closer the instability region is to half-filling.
Extrapolating this behavior we see that the instabilities on the
large-$i$ channels take place only very close to $\mu=0$.

~

The behavior for the extended $s$-wave with a form factor $d(k)=(\cos
k_{x}+\cos k_{y})=\alpha+\beta$ can be studied writing $d(k)$ in
terms of the variables $g$ and $s$ using the solutions of
(\ref{eq:solutions}) and evaluating at $g=0$. We have
$d(s)=\frac{-\mu}{2t}$ and the interaction function reads
\ba
 f(s,s')=U_{0}\;\left(\frac{\mu}{2t}\right)^{2}.
\ea
Again $f(s,s')$ is  independent of the variables $s$ and $s'$ but
now  dependent of the chemical potential $\mu$.

The corresponding instability parameters are obtained by changing
$U_{0}\to U_{0}\left(\frac{\mu}{2t}\right)^{2}$ in
(\ref{eq:chi_for_swave}), and the phase diagram can then be inferred to be analogous to that of Fig. \ref{fig:pd_cte} but with the vertical axes replaced by $U_{0}\left(\frac{\mu}{2t}\right)^{2}$.

%


\subsection{$d$-wave Pomeranchuk instability}
\label{sec:d-wave}

Now we investigate  $d$-wave Fermi Surface Deformation (dFSD)
instability\cite{Scalpino} in the charge channel on a square
lattice. The forward scattering interaction driving the
dFSD has the form \cite{Metzner_PRL}
\ba \label{eq:Q_by_Metzner}
    f(k,k') &=& - G\;d(k)d(k')
\ea
with $G>0$ and $d$-wave form factors $d(k)=(\cos k_{x}-\cos k_{y})$.
The above expression for this effective
interaction was obtained by Metzner {\it et al} \cite{Metzner_PRL} using
functional renormalization group methods.

Using the shorthand notation (\ref{alphabeta}) the interaction reads
\ba
    f(s,s')=-G\;(\alpha-\beta)(\alpha'-\beta')
\ea
and using the  solution of eq. (\ref{eq:solutions}) we have
\ba
      d(s)=\frac{2 }{\cos(2s)}\left( \frac{J^{-1}(s)}{2} - 1  \right)
\ea
Notice that this interaction contains the Jacobian but its origin is
totally independent of the treatment developed in the last sections.

The form factor $d(s)$ can be expanded in a series of the form
\ba\label{eq:desarrollo_d}
    d(s)=\sum_{k=0}^{\infty}d_{k}\cos((4k+2)s)\,,
\ea
where the first coefficients of the expansion are presented in the
Appendix.

Performing the Gram-Schmidt orthogonalization as in previous case, we find the $\chi^\mu$-parameters corresponding to this interaction. The results are very similar, and we will only display here the first two parameters that show an instability of the system, namely
\ba
    \chi_{0}^\mu&=&-2G \,{\pi }^2\,
       d_{0}^2  +
  \pi \,j_{0} +
  \frac{\pi \,j_{1}}{2}\,, \\\nonumber
     \chi_{8}^\mu&=&
     \frac{\pi}{2 g \pi 4{d_{0}}^2 \!-\!
       2 \left( 2 j_{0} \!+\!
       j_{1} \right) }
       \left(
      {j_{1}}^2 \!-4\,{j_{0}}^2 \!+\! 2j_{1}j_{2}+\right.\\\nonumber
       &&+\left.{j_{2}}^2 -j_{1}j_{3} -2j_{0}
       \left( j_{1} +
         j_{3} \right)  + g \pi \left( 2{d_{1}}^2
          \left( 2j_{0} +j_{1} \right)\right.\right.
          \\  && -
          \left.\left.
         8 d_{0}
          d_{1}
          \left( j_{1} +
            j_{2} \right)  +
         4{d_{0}}^2
          \left( 2\,j_{0} +
            j_{3} \right)  \right)
      \right)
\ea

\begin{figure}[t]
  \includegraphics[width=240pt]{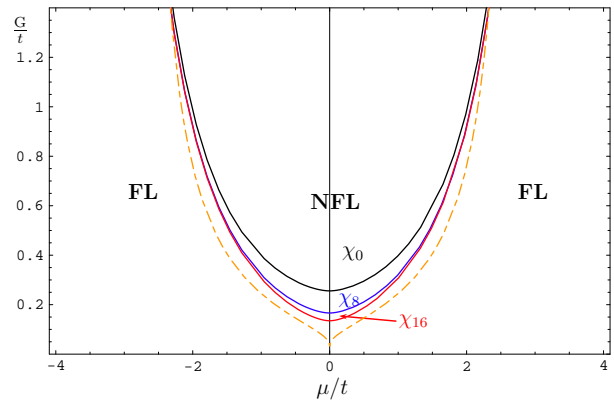}\\
  \caption{Phase diagram for the $d$-wave interaction. The first three unstable channels are shown.
  The dashed line corresponds to the the critical value for the interaction
  parameter studied in reference \onlinecite{nematic_MF} within a Mean Field
  treatment.}\label{graf:phasediag_quarupolar}
\end{figure}

Again, by making use of this first instabilities we can sketch the
phase diagram corresponding to this interaction, as shown in Fig.\
\ref{graf:phasediag_quarupolar}. Similarly to the previous case, as
the interaction grows instabilities appear first in the higher channels.
The dashed line corresponds to the critical value of the interaction
found in ref \onlinecite{nematic_MF} by means of a Mean Field
procedure. Notice that the lowest channels cover most of the
instability zone. The phase diagram shown in Fig.\
\ref{graf:phasediag_quarupolar} is consistent with the results
presented in [\onlinecite{nematic_MF,Kee1}].

Unlike treatments using Mean Field, with the present formalism it is
possible to identify  the region in the parameter space where each
channel  presents a breakdown of the Fermi liquid behavior.

\section{Summary and Conclusions}
\label{sec:conclusions}
In this paper we have developed a general procedure for detecting instabilities in two dimensional lattice models. It is an extension of the formalism of Landau-Pomeranchuk, in particular for lattice systems with an arbitrary shaped FS and allows to describe the phase diagram of the system as an alternative to the usual procedures. The steps are simple and applicable to a wide variety of systems.

Complementarily to other descriptions, our procedure permits to identify the breakdown of the Fermi liquid behavior on each instability channel independently.

As a form of testing our procedure, we have analyzed the stability of the Fermi Liquid in a square lattice, for various interactions already studied in the literature. The $s$-wave  instability in the electronic channel and the instability produced by $d$-wave forward scattering interactions were studied at $T = 0$. The instabilities corresponding to low channels produce a breakdown of the FL behavior for a wide range of fillings, while those occurring for higher channels are closer to half filling. Our result are in good agreement with those obtained by different methods that were previously published by other authors.

Generalization to higher dimensions, spin-dependent interactions or finite temperature can be achieved following the same lines and the results will be presented elsewhere \cite{workinprog}.

\section{Appendix: Orthogonal basis and Series Expansion.}
In this Section we present the coefficients in the expansion of the
functions used in this paper. For the Jacobian, the series takes the
form
 \ba\label{eq_App:desarrollo_j}
    J^{-1}(s)=\sum_{n}\;j^{\mu}_{n}\cos(4ns)\,,
 \ea
where the first coefficients in the expansion are given by
 \ba\label{eq_App:coeff_j2}
 \nonumber    j^{\mu}_{0}&=&\frac{4}
  {\pi }\,\mathbf{E}\left[1 - \frac{{\mu}^2}{16}\right]\\
 \nonumber    j^{\mu}_{1}&=&\frac{1}{\pi}\left(
 |\mu| \,\mathbf{E}\left[1 -
       \frac{16}{{\mu}^2}\right] -
    4\,\mathbf{E}\left[1 - \frac{{\mu}^2}{16}\right]-\right.\\ \nonumber
     &&-\left.
    |\mu| \frac{\,\pi}{2} \;\mathbf{{ _{2}F_{1}}}
    \left[ - \frac{1}{2}  ,\frac{3}{2},2,
        1 - \frac{16}{{\mu}^2}\right]+\right.\\ \nonumber
         &&+\left.
    2\,\pi \;\mathbf{{ _{2}F_{1}}}\left[-
         \frac{1}{2}  ,\frac{3}{2},2,
      1 - \frac{{\mu}^2}{16}\right]\right)\,,\\
 \nonumber    j^{\mu}_{2}&=&\frac{4}{\pi }\,\left( 2\,\mathbf{E}\left[1 -
         \frac{{\mu}^2}{16}\right]+\right.\\ \nonumber
         & &-\left.
      4\,\pi \,\mathbf{{ _{2}F_{1}}}\left[
        - \frac{1}{2}  ,\frac{3}{2},2,
        1 - \frac{{\mu}^2}{16}\right]+\right.\\ \nonumber
        & &+\left.
      3\,\pi \,\mathbf{{ _{2}F_{1}}}\left[
        - \frac{1}{2}  ,\frac{5}{2},3,
        1 - \frac{{\mu}^2}{16}\right] \right)\,, \\
 \nonumber    j^{\mu}_{3}&=&\frac{-8}{\pi }\,\left(\mathbf{E}\left[1 -
       \frac{{\mu}^2}{16}\right]+\right.\\ \nonumber
        &&+
    4\,\pi \,\left( 9\,
        \mathbf{{ _{2}F_{1}}}\left[- \frac{1}
            {2}  ,\frac{3}{2},2,
         1 - \frac{{\mu}^2}{16}\right]-\right.\\ \nonumber
         && -\left.
       18\,\mathbf{{ _{2}F_{1}}}\left[-
             \frac{1}{2}  ,\frac{5}{2},3,
         1 - \frac{{\mu}^2}{16}\right]+\right.\\ \nonumber
         & &+\left.\left.
       10\,\mathbf{{ _{2}F_{1}}}\left[-
             \frac{1}{2}  ,\frac{7}{2},4,
         1 - \frac{{\mu}^2}{16}\right] \right)\right) \,,\\
 \ea
where $\mathbf{E}\left[m\right]$ is the complete elliptic integral
 \ba
    \mathbf{E}\left[ m \right]=\int_{0}^{\frac{\pi}{2}}\sqrt{1-m\sin^{2}(t)}\,\;dt\,,
 \ea
and \  $\mathbf{{ _{2}F_{1}}}(a,b;c;z)$  is the hypergeometric
function
\ba
 \mathbf{{ _{2}F_{1}}}(a,b;c;z)&=&\frac{\Gamma(c)}
  {\Gamma(b)\,\Gamma(-b +
  c)}\times\\ \nonumber
  &&\times \int_{0}^{1}t^{b-1}(1-t)^{c-b-1}(1-t z)^{-a}\;dt\,.
\ea

The form factor  for the d-wave forward scattering interaction can
be expanded as follows
\ba\label{eq_App:desarrollo_d}
    d(s)=\sum_{k=0}^{\infty}d_{k}\cos((4k+2)s)\,,
\ea
with
 \ba\label{eq_App:coeff_j}
 \nonumber    d^{\mu}_{0}&=&\frac{-4\,\left( \pi  -
      2\,\mathbf{E}\left[1 - \frac{{\mu}^2}{16}\right]
      \right) }{\pi }\\
 \nonumber    d^{\mu}_{1}&=&\frac{4}{\pi }\,\left( \pi  -
      6\,\mathbf{E}\left[1 -
         \frac{{\mu}^2}{16}\right]+\right.\\ \nonumber
          &&+\left.
      2\,\pi \,\mathbf{{ _{2}F_{1}}}\left[
        - \frac{1}{2}  ,\frac{3}{2},2,
        1 - \frac{{\mu}^2}{16}\right] \right) \,,\\
 \nonumber    d^{\mu}_{2}&=&\frac{-4}{\pi }\,\left( \pi  -
      10\,\mathbf{E}\left[1 -
         \frac{{\mu}^2}{16}\right]+\right.\\\nonumber
          &&+\left.
      10\,\pi \,\mathbf{{ _{2}F_{1}}}\left[
        - \frac{1}{2}  ,\frac{3}{2},2,
        1 - \frac{{\mu}^2}{16}\right]-\right.\\\nonumber
        && -\left.
      6\,\pi \,\mathbf{{ _{2}F_{1}}}\left[
        - \frac{1}{2}  ,\frac{5}{2},3,
        1 - \frac{{\mu}^2}{16}\right] \right) \,,\\
 \nonumber    d^{\mu}_{3}&=&\frac{4}
    {\pi }\,\left( -14\,\mathbf{E}\left[1 -
         \frac{{\mu}^2}{16}\right]+\right.\\\nonumber
         &&+\left.
      \pi \,\left( 1 +
         28\,\mathbf{{ _{2}F_{1}}}\left[-
              \frac{1}{2}  ,\frac{3}{2},2,
           1 - \frac{{\mu}^2}{16}\right]-\right.\right.\\\nonumber
           &&-\left.\left.
         42\,\mathbf{{ _{2}F_{1}}}\left[-
              \frac{1}{2}  ,\frac{5}{2},3,
           1 - \frac{{\mu}^2}{16}\right]+\right.\right.\\\nonumber
            &&+\left.\left.
         20\,\mathbf{{ _{2}F_{1}}}\left[-
              \frac{1}{2}  ,\frac{7}{2},4,
           1 - \frac{{\mu}^2}{16}\right] \right) \right)\,.
 \ea

The orthogonal basis $\{\xi_i\}$ depends of course on the specific form of the interaction, but
in all the cases studied here it satisfy the following properties:

1 - The functions are either linear combinations of $\sin(s)$ or of $\cos(s)$
separately. There is not mixtures of $\sin$ and $\cos$.

2 - All the functions reduce to the expressions corresponding to the free case in
the limit when the interaction parameter is sent to zero.

\section*{ACKNOWLEDGMENTS: }
We would like to thank  E. Fradkin for helpful discussions. This
work was partially supported by the ESF grant INSTANS, ECOS-Sud
Argentina-France collaboration (Grant No A04E03), PICS CNRS-Conicet
(Grant No. 18294), PICT ANCYPT (Grant No 20350), and PIP CONICET
(Grant No 5037).

%

\end{document}